\documentclass[sigconf,natbib=true]{acmart} %
\usepackage{graphicx,verbatimbox}
\usepackage{listings}
\usepackage{amsmath}
\usepackage{subcaption}
\usepackage{mathtools}
\usepackage{color}
\usepackage{tabularx,ragged2e}
\usepackage{multirow}
\usepackage{mathtools}
\usepackage{enumitem}
\usepackage[T2A,T1]{fontenc}
\usepackage[utf8]{inputenc}
\usepackage[russian,english]{babel}
\usepackage{xcolor}
\usepackage{booktabs}

\graphicspath{{images/}}

\AtBeginDocument{%
	\providecommand\BibTeX{{%
			\normalfont B\kern-0.5em{\scshape i\kern-0.25em b}\kern-0.8em\TeX}}}

\copyrightyear{2023}
\acmYear{2023}
\acmConference[Gen-IR@SIGIR2023]{The First Workshop on Generative Information Retrieval}{July 27, 2023}{Taipei, Taiwan}
\acmBooktitle{The First Workshop on Generative Information Retrieval (Gen-IR@SIGIR23), July 27, 2023, Taipei, Taiwan}
\acmDOI{}
\acmPrice{}
\acmISBN{}
\setcopyright{rightsretained}

\begin{document}
	

\title[Bridging the Gap Between Indexing and Retrieval for
Differentiable Search Index with Query Generation]{Bridging the Gap Between Indexing and Retrieval for
	Differentiable Search Index with Query Generation}




\author{Shengyao Zhuang}
\affiliation{%
	\institution{The University of Queensland}
	\country{Australia}}
\email{s.zhuang@uq.edu.au}
\authornote{This work was done during an internship at Microsoft STCA}


\author{Houxing Ren}
\affiliation{%
	\institution{Beihang University}
	\country{China}}
\email{renhouxing@buaa.edu.cn}
\authornotemark[1]

\author{Linjun Shou}
\affiliation{%
	\institution{Microsoft STCA}
	\country{China}}
\email{lisho@microsoft.com}

\author{Jian Pei}
\affiliation{%
	\institution{Simon Fraser University}
	\country{Canada}}
\email{jpei@cs.sfu.ca}

\author{Ming Gong}
\affiliation{%
	\institution{Microsoft STCA}
	\country{China}}
\email{migon@microsoft.com}

\author{Guido Zuccon}
\affiliation{%
	\institution{The University of Queensland}
	\country{Australia}}
\email{g.zuccon@uq.edu.au}
\authornote{Corresponding author.}

\author{Daxin Jiang}
\affiliation{%
	\institution{Microsoft STCA}
	\country{China}}
\email{djiang@microsoft.com}
\authornotemark[2]


\begin{abstract}
	The Differentiable Search Index (DSI) is an emerging paradigm for information retrieval. Unlike traditional retrieval architectures where indexing and retrieval are two different and separate components, DSI uses a single transformer model to perform both indexing and retrieval. 
	
	In this paper, we identify and tackle an important issue of current DSI models: the data distribution mismatch that occurs between the DSI indexing and retrieval processes. Specifically, we argue that, at indexing, current DSI methods learn to build connections between the text of long documents and the identifier of the documents, but then retrieval of document identifiers is based on queries that are commonly much shorter than the indexed documents.
	 This problem is further exacerbated when using DSI for cross-lingual retrieval, where document text and query text are in different languages. 
	 
	 To address this fundamental problem of current DSI models, we propose a simple yet effective indexing framework for DSI, called DSI-QG. When indexing, DSI-QG represents documents with a number of potentially relevant queries generated by a query generation model and re-ranked and filtered by a cross-encoder ranker. The presence of these queries at indexing allows the DSI models to connect a document identifier to a set of queries, hence mitigating data distribution mismatches present between the indexing and the retrieval phases. Empirical results on popular mono-lingual and cross-lingual passage retrieval datasets show that DSI-QG significantly outperforms the original DSI model.

\end{abstract}

%

\keywords{Differentiable search index, query generation, retrieval via autoregressive generation}

\maketitle

\section{Introduction} \label{intro}

\begin{figure*}
    \centering
    \includegraphics[width=1\linewidth]{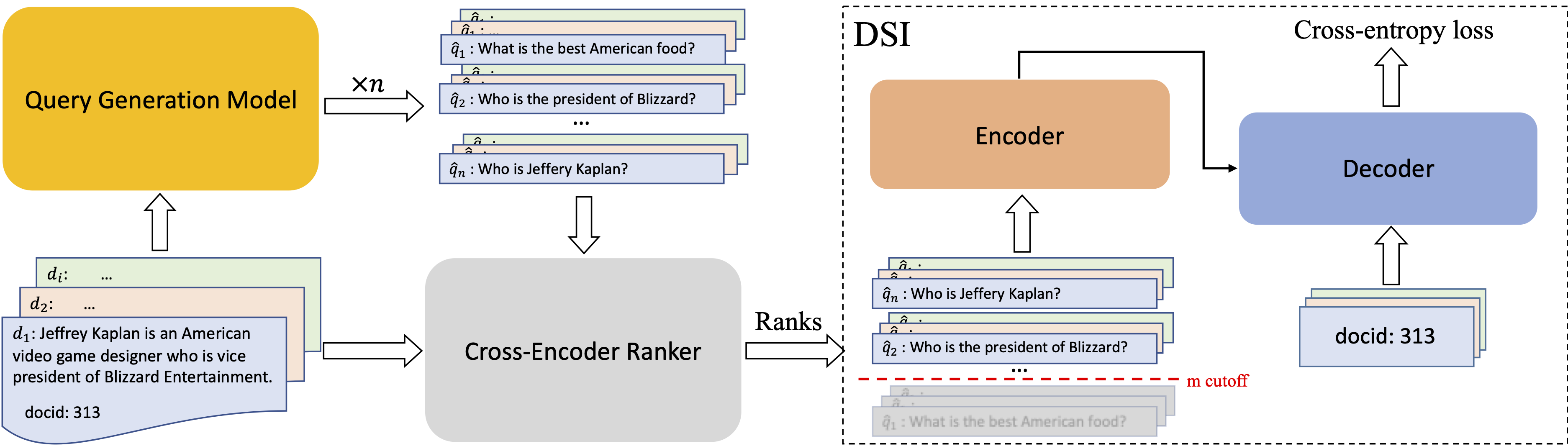}
    \caption{Overview of the proposed DSI-QG indexing framework, where a query generator (yellow box) is used to generate potential queries, which in turn are used to represent documents for indexing. The architecture of the method includes a cross-encoder ranker (gray box) that is used to select only promising queries to be included during indexing.}
    \label{fig:pipeline}
\end{figure*}

Information retrieval (IR) systems aim to return a ranked list of relevant documents for a given user query. Most modern information retrieval systems are based on the index-then-retrieve pipeline where documents are first encoded and stored in an inverted index~\cite{zobel2006inverted,formal2021splade,gao-etal-2021-coil,lin2021few,zhuang2021fast,dai2020context,mallia2021learning} or a nearest neighbor search index~\cite{chappell2015approximate,lin2022proposed,karpukhin-etal-2020-dense,xiong2020approximate,gao2022unsupervised,zhan2021optimizing,lin2020distilling,khattab2020colbert} and search results are then constructed based on a retrieval model that exploits the information in the index. By doing so, the indexing and retrieval processes are decoupled.

Recently, an alternative approach called Differentiable Search Index (DSI) has been proposed~\cite{DBLP:journals/corr/abs-2202-06991}. Instead of separating indexing and retrieval into two different components in an IR system, DSI aims to encode all information of the corpus and conduct retrieval within a single Transformer language model~\cite{vaswani2017attention}. To do so, in the indexing phase, DSI learns to build connections within its model parameters between the text in documents and the corresponding document identifiers (docids). Then, in the retrieval phase, the trained Transformer model takes as input a query text and directly outputs a potentially-relevant ranked docid using beam search. Compared to traditional IR pipelines, DSI learns an end-to-end search system in a unified manner, thus considerably simplifying the architecture of IR systems.

Despite the original DSI method being shown  effective on the document retrieval task~\cite{DBLP:journals/corr/abs-2202-06991}, in this paper we argue that this model is affected by a significant data distribution mismatch problem. More specifically, in the indexing phase, long text from documents is fed as input to the DSI model. However, in the retrieval phase, the model only observes short query texts as inputs. Therefore, the input data at indexing time is considerably different from the input data at retrieval time. It is well-known that pre-trained language models are not robust to data distribution drift between training (indexing) and inference (retrieval)~\cite{li2020bert,ma2020charbert,li2021searching,wang-etal-2021-textflint,zhuang2021dealing,zhuang2022characterbert}: we thus argue that the original DSI model might be sub-optimal. Furthermore, in our experiments we demonstrate that the negative impact of the data distribution mismatch problem is even more considerable when adapting DSI to the cross-lingual retrieval setting, where documents and queries are from different languages.

Intuitively, DSI may be more effective for collections of short documents because short documents are similar to queries -- at least in terms of text length. Thus, the data distribution mismatch problem may be lesser when documents are short in length. 
Indeed, in the original paper, although not explicitly recognising the data distribution mismatch problem, \citeauthor{DBLP:journals/corr/abs-2202-06991} have shown that truncating long documents into shorter lengths displays higher effectiveness\cite[Section 4.4.5]{DBLP:journals/corr/abs-2202-06991}. To further improve effectiveness, they also attempted adding labeled relevant queries into the indexing data so that the DSI model could learn to directly link a query to its relevant document identifier. However, for the majority of the documents in the collection where there is no labeled query provided, the model builds connections for the document identifiers with the original document texts only, as it is never exposed to the corresponding relevant queries: hence the data distribution mismatch problem still persists.

Based on our intuition of the data distribution mismatch problem that affects the DSI model, in this paper we propose \textit{DSI-QG}, a simple yet effective indexing framework for DSI. The core idea of DSI-QG is that, instead of using the original long text from documents for indexing, DSI-QG uses a set of queries that are relevant to the original document for indexing. Specifically, for each document in the corpus, we employ a query generation model to generate a large set of potentially relevant queries which we use to represent each document. For the cross-lingual retrieval task, this query generation model is trained to be able to generate queries in different languages. To control the quality of the generated queries, all the generated queries are fed into a cross-encoder ranker along with their corresponding documents. This model ranks all the generated queries according to their relevance to the document; then queries are filtered to only pass the top-$m$ most relevant queries to the DSI module for indexing.
By doing so, the same type of data is fed into the DSI in both the indexing and retrieval phases, hence avoiding the data distribution mismatch problem. Figure~\ref{fig:pipeline} illustrates our proposed DSI-QG indexing framework.

Our contributions can be summarised as follows:
\begin{itemize}[noitemsep,nolistsep,leftmargin=*]
\item We identify a crucial problem that affects the original DSI model: the data distribution mismatch between indexing and retrieval.
\item We show that DSI performs poorly in the presence of data distribution mismatches: this is further exacerbated in the cross-lingual document retrieval setting, emphasizing the gap between documents and queries. 
\item We propose the DSI-QG indexing framework which is aimed at tackling the data distribution mismatch problem. Our framework uses query generation models (including a cross-lingual query generation model) and a cross-encoder ranker to generate and rank a set of potentially relevant queries that are used to represent documents for indexing. 
\item We conduct extensive experiments on both mono-lingual and cross-lingual document retrieval datasets. Our results show that, with our proposed framework, the effectiveness of DSI is improved by a large margin on these tasks. 
\end{itemize}
\vspace{2pt}

Code to reproduce the experiments and results presented in this paper can be found at \url{https://github.com/ArvinZhuang/DSI-QG}.

\section{Preliminaries} \label{preliminaries}
In this section, we describe the details of the original DSI method. Then in the next section, we introduce our DSI-QG framework.

DSI performs index-then-retrieve with a single T5 transformer model~\cite{raffel2019exploring}. During the indexing phase, the DSI model is fine-tuned to associate the text string of each document $d_i$ in a collection $D$ with its corresponding document identifier (docid) $i$. It utilizes a straightforward sequence-to-sequence (seq2seq) approach that takes the document text as input and generates docids as output~\cite{nogueira2019doc2query}. The model is trained with the standard T5 training objective which uses the teacher forcing policy~\cite{williams1989learning} and the cross-entropy loss:
\begin{equation}\label{eq1}
\mathcal{L}_{DSI}(\theta) = \sum_{d_i \in D}\log p{(i | T5_{\theta}(d_i))}
\end{equation}

The docid $i$ can be represented using a single token (\textit{Atomic Docid}) or a string of tokens (\textit{String Docid})~\cite{DBLP:journals/corr/abs-2202-06991}. For the Atomic Docid, each docid is a single token in the T5 vocabulary and it has been encoded as an embedding vector in the T5 embedding layer. Thus the task can be considered as an extreme multi-label classification problem~\cite{liu2021emerging} where the model  learns a probability distribution over the docid embeddings. This setting poses a limit to DSI when used on large-scale corpora, since the size of the T5 embedding layer cannot be too large. Hence, we do not consider this setting in our experiments.

On the other hand, the String Docid strategy treats the docids as arbitrary strings so that they can be generated in a token-by-token manner with the original T5 vocabulary. This configuration does not pose limitations with respect to corpus size. The original DSI paper also proposed a \textit{Semantic String Docid} which uses a hierarchical clustering algorithm to force similar documents to have similar docids. Since clustering docids is not the aim of this paper and for simplicity, we only consider arbitrary String Docid, and leave extensions to the Semantic String Docid setting to future work.

In the retrieval phase, given an input query $q$, a DSI model returns a docid by autoregressively generating the docid string using the fine-tuned T5 model. The probability of the generated docid $i$ can be computed as:
\begin{equation}\label{eq2}
p(i|q, \theta) = \prod_{t=1} p(i_t | T5_{\theta}(q,i_0, i_1,..., i_{t-1})),
\end{equation}
where $i_t$ is the $t$-th token in the docid string. A ranked list of docids is then constructed using beam search (and is thus ranked by decreasing generation probability).

It is important to note that a query usually is much shorter in length than a document. This means the length of the input data at indexing is very different from the length of input data at retrieval: thus the DSI model suffers from a data distribution mismatch. To mitigate this problem, the DSI authors proposed the use of the supervised DSI fine-tuning~\cite{DBLP:journals/corr/abs-2202-06991}. This process adds labeled relevant queries to the indexing data. Let $Q_i$ be the set of labeled relevant queries for $d_i$, then the training objective becomes:
\begin{equation}
\begin{split}
\mathcal{L}_{DSI-S}(\theta) = &\sum_{d_i \in D} \log p{(i | T5_{\theta}(d_i))} + \\ &\sum_{q_j \in Q_i}\log p{(i | T5_{\theta}(q_j))},
\end{split}
\end{equation}
where $q_j \in Q_i$ is a query relevant to $d_i$. We note that having assessors labeling a relevant query for all the documents in the collection requires a large annotation effort thus not all documents can have a human-judged relevant query for supporting indexing. In other words, $Q_i$ could be an empty (or small) set. Hence the DSI model still largely suffers from the data distribution mismatch problem (especially for large collections of documents), even when the supervised DSI fine-tuning method is used.

\section{The DSI-QG Framework} \label{method}
In this section, we describe the details of the different components in the proposed DSI-QG framework, pictured in Figure~\ref{fig:pipeline}. Our framework features a query generation model for generating candidate queries that are potentially relevant to the original documents. It then uses a cross-encoder ranker to rank all generated queries and only selects the top-$n$ queries which are then passed to the downstream DSI module for representing the associated document at indexing.

\subsection{DSI with query generation}
The original DSI method exhibits a gap between the input data used at indexing and that used at retrieval.  In order to bridge this gap and improve DSI's effectiveness, we propose DSI-QG which uses a query generation model to generate a set of potentially-relevant queries to represent each candidate document for indexing.  Specifically, we denote $\hat{Q_i}$ as the set of queries generated by a query generation model $QG$ given the document $d_i$:
\begin{equation}
\hat{Q_i} = QG(d_i).
\end{equation}

All the generated queries $\hat{q}$ in $\hat{Q_i}$ share the same docid as $d_i$, and $|\hat{Q_i}|=n$.
We then replace the original documents that need to be indexed with their corresponding generated queries, i.e. using $\hat{Q_i}$ in place of $d_i$. In other words, a document is replaced by the set of queries generated for that document.
Thus, during the indexing phase in DSI-QG, the modified DSI model is trained to associate the generated queries of each candidate document with its docid:
\begin{equation}
	\mathcal{L}_{DSI-QG}(\theta) = \sum_{d_i \in D} \sum_{\hat{q_j} \in \hat{Q_i}}\log p{(i | T5_{\theta}(\hat{q_j}))}.
\end{equation}

The retrieval phase of DSI-QG is the same as the original DSI model and takes a user query as input and uses beam search to construct a ranked docid list. Note that each query in $\hat{Q_i}$, that was generated for $d_i$, is used separately for the other queries for $d_i$, i.e. queries for a document are not concatenated or combined into a single input. 
In summary, in our DSI-QG framework, a DSI model only observes short queries as input data during both indexing and retrieval thus eliminating the data distribution mismatch problem that affects the original DSI model.

A key factor for the success of the DSI-QG indexing framework is the query generation (QG) model. This model should generate high-quality and diverse relevant queries so that they can effectively represent the corresponding document from which they are generated. For this purpose, we train a T5 model with a similar seq2seq objective as Eq.~\eqref{eq1}, but in this case the input is the document text and the output is the labeled relevant query $q_j$:
\begin{equation}
	\mathcal{L}_{QG}(\theta) = \sum_{d_i \in D}\sum_{q_j \in Q_i}\log p{(q_j | T5_{\theta}(d_i))}.
\end{equation}

After training a QG model, instead of using beam search, we use a random sampling strategy to generate a set of queries for each candidate document.  This is because we find random sampling gives more creative and diverse queries than beam search, thus potentially covering more relevant information about the document. To avoid randomly generating irrelevant content and grammatically incorrect queries, we adopt the top-$k$ sampling scheme~\cite{fan2018hierarchical} which only allows the $k$ most likely next words to be sampled during the query generation and the probability mass of those $k$  next words is redistributed. In our experiments, we set $k=10$. 

Intuitively, a document may be relevant to more than just one query, thus another factor that might impact the effectiveness of our proposed DSI-QG method is the number of generated queries $n$ to represent each document: we discuss the impact of $n$ in the result section.

\subsection{DSI-QG with cross-lingual query generation}
To generalize our DSI-QG framework to the cross-lingual IR setting, we also train a multi-lingual T5 model~\cite{xue2021mt5} to generate queries in different languages, and then in turn use these to represent a document. To achieve this, we use a prompt-based approach to control the generated query language. Specifically, we place the target language and the document text in the following template for both training and inference:
\begin{equation}
\texttt{Generate[lang]question:[doc]},
\end{equation}
where \texttt{[lang]} and \texttt{[doc]} is the placeholder for the target query language and the document text. In our cross-lingual experiments, \texttt{[doc]} is always written in English and \texttt{[lang]} is a language other than English. We generate multiple queries for all the target languages and use these to represent each English document. By doing so, our DSI-QG model can learn to build connections between the English documents identifiers with queries from different languages, thus allowing to perform cross-lingual retrieval with our proposed cross-lingual query generation model.

\subsection{Ranking generated queries with a cross-encoder ranker}
Although our query generation model adopts the top-$k$ sampling scheme to balance the relevance and diversity of generated queries, it still inevitably generates irrelevant queries due to the randomness of the sampling process. This problem is even more considerable when there is not enough training data to train the query generation model or when the model is ill trained. To further mitigate this problem, we add a cross-encoder ranker to rank all the generated queries and only use the top-$m$ ranked queries to represent the original document.

Specifically, we use monoBERT~\cite{nogueira2019multi} as a cross-encoder ranker: this is a transformer encoder-based model that employs BERT and that takes a query-document pair (separated by a special \texttt{[SEP]} token) as input and outputs a relevance score $s$:

\begin{equation}
	Ranker(q, d) = BERT(\texttt{[q][SEP][d]}) = s,
\end{equation}

We train the ranker with supervised contrastive loss, similar to \citeauthor{gao2021rethink}~\cite{gao2021rethink}:

\begin{equation}
	\mathcal{L}_{Ranker} = \sum_{q\in Q}-\log \frac{e^{Ranker(q, d^+)}}{e^{Ranker(q, d^+)} + \sum_{d^-}e^{Ranker(q, d^-)}},
\end{equation}

\noindent where $q $ is the training query and $d^+$ is the annotated relevant document for the training query. $d^-$ is a hard negative document which we randomly sample from the top 100 documents retrieved by BM25 for the training query $q$.

In our cross-encoder ranker, all the query tokens can interact with all the document tokens thus it has more substantial relevance modeling power than other ranker architectures, such as dual- or bi-encoders~\cite{lin2020distilling,ren2021rocketqav2,zhang2022adversarial,lu2022ernie}.  We then rank all the generated queries for each document in decreasing order of the relevance score estimated by our ranker. From this ranking, we only select the top-$m$ queries to pass to the downstream DSI indexing training, thus effectively filtering out the remaining $n-m$ queries. We note that our query generation model and cross-encoder ranker are large transformer models and thus require substantial computational resources in addition to the DSI model alone. However, these additional computations only happen during the offline indexing phase, and will not affect the online query latency. We leave methods for reducing the computational resources required for indexing to future work.
\section{Experimental Settings}

\subsection{Datasets}
Following the original DSI paper, we conduct our experiments on subsets of publicly available document retrieval datasets, namely NQ 320k~\cite{kwiatkowski2019natural}, for the mono-lingual document retrieval task, and XOR QA 100k~\cite{asai2021xor}, for the cross-lingual retrieval task. The NQ 320k dataset has $\approx$307k training query-document pairs and $\approx$8k dev query-document pairs. All the queries and documents in NQ 320k are in English. We follow the description in DSI~\cite{DBLP:journals/corr/abs-2202-06991} and SEAL~\cite{bevilacqua2022autoregressive} to construct the dataset as the code for dataset construction is not yet publicly available at the time of writing. For XOR QA 100k, we use the gold paragraph data available in the original repository\footnote{\url{https://github.com/AkariAsai/XORQA\#gold-paragraph-data}} which contains around 15k gold (annotated as relevant) document-query pairs in the training set and 2k gold document-query pairs in the dev set. Queries in both train and dev sets are in 7 typologically diverse languages\footnote{They are Ar, Bn, Fi, Ja, Ko, Ru and Te.} and documents are in English. The total number of documents in the XOR QA training set and dev set is around 17k. This is a very small number of documents, likely to render the retrieval task too easy. We then randomly sample 93k documents from a dump of the English Wikipedia corpus to form a 100k collection for testing our models, thus increasing how challenging retrieval in this collection is.

\subsection{Baselines}

We compare DSI-QG with the following baselines:

\begin{itemize}[]
\item BM25~\cite{robertson2009probabilistic}: a classic sparse retrieval method based on inverted indexes. This method usually only works for mono-lingual retrieval tasks as it is a keyword-matching  method. We use the Pyserini~\cite{Lin_etal_SIGIR2021_Pyserini} implementation of BM25 for this baseline. 
\item BM25 + docT5query~\cite{nogueira2019doc2query}: a sparse retrieval method which also leverages query generation. It uses a T5 model to generate a set of queries and appends them to the original document. Then it uses an inverted index and BM25 to retrieve augmented documents. In the original study that investigated this method, only the mono-lingual retrieval task was considered~\cite{nogueira2019doc2query}. For fair comparison with DSI-QG, we adapt this method to the cross-lingual retrieval setting by replacing the mono-lingual T5 query generation model with the same multi-lingual T5 generation model used in our DSI-QG.  
We also use the Pyserini implementation for this baseline.
\item SEAL~\cite{bevilacqua2022autoregressive}: an autoregressive generation model that is similar to DSI. It treats ngrams that appear in the collection as document identifiers; at retrieval time, it directly generates and scores distinctive ngrams that are mapped to the documents. Unlike DSI, which unifies the index into the model parameters, SEAL requires a separate index data structure to perform an efficient search. Note that no publicly available implementation of SEAL currently exists. Unlike for DSI below, the re-implementation of SEAL is outside the scope of our work, and thus we report the results obtained by Bevilacqua et al.~\cite{bevilacqua2022autoregressive} on the NQ 320k dataset. SEAL has not been devised for and experimented with the task of cross-lingual retrieval and thus no results for XOR QA 100k are reported.

\item mDPR~\cite{karpukhin-etal-2020-dense,asai2021xor}: a mBERT-based cross-lingual dense passage retrieval method trained with a contrastive loss and with hard negatives sampled from the top passages retrieved by BM25. mDPR relays on nearest neighbor index search  (Faiss implementation~\cite{johnson2019billion}) to retrieve the passages that have the closest embeddings to the query embedding. 
We train the mDPR model with the Tevatron dense retriever training toolkit~\cite{Gao2022TevatronAE}. Of course, due to its cross-lingual nature, we run mDPR only on the cross-lingual dataset, XOR QA 100k.
\item DSI~\cite{DBLP:journals/corr/abs-2202-06991}: The original DSI method that uses documents text as input for indexing. Since the original code has not currently been made available by the authors, we implement and train the DSI model ourselves using the Huggingface transformers Python Library. We provide the implementation of this DSI model in our public code repository, along with the implementations of the other models considered in this paper.
\end{itemize}

\subsection{Evaluation Measures}

Following the original DSI paper, for both datasets, we evaluate baselines and our models on the dev set with Hits@1 and Hits@10. This metric reports the proportion of the correct docids ranked in the top 1 and top 10 predictions. In addition, for XOR QA 100k we also report nDCG@10; this metric is not available for NQ 320k for some of the considered baselines and thus we do not report it as comparisons between methods cannot then be made. 

\begin{table}[b]
		\captionof{table}{Experimental results on NQ 320k datasets. BM25 + docT5query and DSI-QG use the top $m=50$ re-ranked generated queries. 
		}
	\resizebox{0.9\columnwidth}{!}{
		\begin{tabular}{l|cc}
			\hline
			\multirow{2}{*}{Model} & \multicolumn{2}{c}{NQ 320k}                              \\ \cline{2-3} 
			& \multicolumn{1}{c}{Hits@1} & \multicolumn{1}{c}{Hits@10} \\ \hline
			BM25                   & 29.27                      & 60.15                       \\
			BM25 + docT5query      & 39.13                      & 69.72                       \\
			SEAL                   & 26.30                      & 74.50                       \\
			DSI-base               & 27.40                      & 56.60                       \\
			DSI-large              & 35.60                      & 62.60                       \\ \hline
			DSI-QG-base            & 63.49                & 82.36                  \\
			DSI-QG-large           & \textbf{65.13}                      & \textbf{82.50}                       \\ \hline
		\end{tabular}
	}
\label{table1}

\end{table}

\subsection{Implementation Details}
There are three Transformer models in our DSI-QG framework: a query generation model, a cross-encoder ranker, and a DSI model. 

For the NQ 320k dataset, we fine-tune an existing docT5query query generation model checkpoint\footnote{\url{https://huggingface.co/castorini/doc2query-t5-large-msmarco}} with the training portion of the NQ 320k dataset. For the cross-encoder ranker, we train a `BERT-large-uncased' checkpoint and 15 hard negatives documents sampled from BM25. For the DSI model, we use the standard pre-trained T5 model~\cite{raffel2019exploring} to initialize the model parameters. 

For XOR QA 100k, we use the multi-lingual T5 model~\cite{xue2021mt5} to initialize both the query generation model and DSI model. For the cross-lingual ranker, we train `xlm-roberta-large'~\cite{conneau2020unsupervised} checkpoint with BM25 hard negatives provided by the XOR QA official repository. For our trained query generation model, we train the model with a batch size of 128 and a learning rate of $1e^{-4}$ with Adam optimizer for 600 training steps on XOR QA 100k datasets,which is equivalent to about 6 epochs, and 9500 steps on the NQ 320k dataset which is equivalent to about 4 epochs. The DSI models in our DSI-QG method are trained for a maximum of 1M steps with a batch size of 256 and a learning rate of $5e^{-5}$ with 100k warmup-steps. Since the documents in DSI-QG are represented by generated short queries, we set the maximum length of the input data to 32 tokens for faster training and saving GPU memory usage. For training the original DSI model, we use the training configuration suggested in the original paper~\cite{DBLP:journals/corr/abs-2202-06991}. For mDPR trained on XOR QA, we follow the training configuration in the XOR QA paper~\cite{asai2021xor}, which uses a multi-lingual BERT-base model as the backbone query and passage encoder. All Transformer models used in this paper are implemented with Huggingface transformers~\cite{wolf-etal-2020-transformers} and training is conducted on 8 Tesla A100 GPUs.

\section{Results} \label{result}
\subsection{Effectiveness on Mono-lingual Retrieval}


\begin{table*} []
	\centering
		\caption{Experimental results on XOR QA 100k datasets. BM25 + docT5query and DSI-QG use 70 re-ranked generated queries (10 for each language). Improvements for DSI-QG that are statistically significantly better/worse than mDPR are labelled with  $^\star$ ($p<0.05$) and $^\diamond $ ($p<0.01$). DSI-QG methods are always statistically significantly better than the remaining baselines (with $p<0.01$). Statistical analysis performed using two-tailed paired t-test with Bonferroni correction.}
	\begin{subtable}[c]{1\linewidth}
		\centering
		\scalebox{1.2}{
			\begin{tabular}{l|ccccccc|c}
				\hline
				Model          & Ar    & Bn    & Fi    & Ja    & Ko    & Ru    & Te    & Average \\ \hline
				BM25 + docT5query & 11.96 & 19.21 & 29.17 & 20.83 & 10.21 & 25.96 &  8.02 & 17.91   \\
				mDPR             & 20.93 & 19.21 & 43.59 & 22.50 & 20.07 & 41.70 & 18.99 & 26.71   \\
				DSI-base         &    0.00   &  0.00     &   1.28    &     0.00  &   0.70    &  0.00     &   0.00  & 0.28  \\
				DSI-large        &    0.33   &    0.99   &    6.41   &    1.25   &    0.00   &   1.27    &   0.00    & 1.47  \\ \hline
				DSI-QG-base      & 34.55$^\diamond $  & 38.41$^\diamond $ & 42.95 & 42.08$^\diamond $ & \textbf{33.80} $^\diamond $ & 57.45$^\diamond $  & 28.69$^\star $  & 39.17$^\diamond $    \\
				DSI-QG-large     & \textbf{37.21}$^\diamond $  & \textbf{43.05}$^\diamond $ & \textbf{45.19} & \textbf{43.33}$^\diamond $  & 32.04$^\diamond $  & \textbf{61.28}$^\diamond $ & \textbf{31.22}$^\diamond $ & \textbf{41.90}$^\diamond $ \\ \hline
			\end{tabular}
		}
		\caption{Hits@1}
		\vspace{2mm}           
	\end{subtable}
	
	\quad%
	\begin{subtable}[c]{1\linewidth}
		\centering
		\scalebox{1.23}{
			\begin{tabular}{l|ccccccc|c}
				\hline
				Model          & Ar    & Bn    & Fi    & Ja    & Ko    & Ru    & Te   & Average \\ \hline
				BM25 + docT5query & 28.24 & 37.75 & 46.47 & 41.67 & 23.59 & 40.43 & 25.32 & 34.78   \\
				mDPR             & 56.48 & 61.59 & \textbf{73.40} & 50.83 & 53.52 & 72.34 & 54.85 & 60.43   \\
				DSI-base         &   1.66   &   2.65    &   7.05    &    2.91   &    2.47   &    0.85   &   1.26  & 2.69  \\
				DSI-large        &    3.99   &    5.23   &    16.67   &   5.83    &    4.93   &   4.25   &    2.53   & 6.21  \\ \hline
				DSI-QG-base      & \textbf{59.14} & 68.54 & 68.27 & 64.58$^\diamond $ & \textbf{61.97}$^\star $  & 71.91 & \textbf{67.09}$^\diamond $  & 65.93$^\diamond $   \\
				DSI-QG-large     & 58.47 & \textbf{73.18}$^\diamond $  & 73.08 & \textbf{67.08}$^\diamond $  & 59.51 & \textbf{74.04} & 63.71$^\star $ & \textbf{67.01}$^\diamond $   \\ \hline
			\end{tabular}
		}
		\caption{Hits@10}    
		\vspace{2mm}      
	\end{subtable}
	\quad%
\begin{subtable}[c]{1\linewidth}
	\centering
	\scalebox{1.3}{
		\begin{tabular}{l|ccccccc|c}
			\hline
			Model                          & Ar    & Bn       & Fi       & Ja       & Ko    & Ru       & Te       & Average \\ \hline
			BM25 + docT5query & 9.83 & 12.82 & 14.52 & 13.66 & 7.40 & 13.61 & 9.30 & 11.59   \\
			mDPR                          & \textbf{20.22} & 21.05 & 23.0 & 17.84 & 18.37& \textbf{23.43} & 19.97 & 20.55  \\
			DSI-base                     &   0.32   &  0.75    &   1.96    &    0.25  &    0.65   &    0.58   &   0.27  & 0.73 \\
			DSI-large                     &    1.57   &    2.06   &    5.14   &   0.20   &    2.23   &   0.90   &    0.93   & 2.12  \\ \hline
			DSI-QG-base              & 19.28 & 21.47 & 20.79 & 20.88 & \textbf{19.64}  & 21.94 & \textbf{22.59}  & 20.94   \\
			DSI-QG-large     & 18.80 & \textbf{22.86} & \textbf{23.54} & \textbf{21.09}  & 19.32& 22.24& 20.47& \textbf{21.19}   \\ \hline
		\end{tabular}
	}
	\caption{nDCG@10}    
\end{subtable}
\label{table2}
\end{table*}

We start by discussing the effectiveness of the proposed DSI-QG framework on the mono-lingual  retrieval task; recall that these experiments are based on the NQ 320k English mono-lingual retrieval dataset.

Table~\ref{table1} contains the Hits scores of the baselines and our DSI-QG methods on NQ 320k. For BM25 + docTquery and DSI-QG, we first generated $n=100$ queries for each document; then we ranked them using the cross-encoder ranker and select only the top $m=50$ queries. This process thus resulted in 50 queries being used to represent each document. To explore the impact of different model sizes, we report the results for DSI and DSI-QG with T5-base (200M parameters) and T5-large (800M). 

The results show that the original DSI method performs worse than other baselines, with the exception of DSI with T5 large which outperforms BM25 on both Hits scores and SEAL on Hits@1. BM25 with docT5query document augmentation, which is a simple and straightforward way of leveraging query generation, achieves the best Hits@1 among the baselines we consider. These results suggest that the existing autoregressive generation-based information retrieval methods are inferior to the considered baselines in the mono-lingual retrieval task. 

On the other hand, our DSI-QG outperforms all baselines by a large margin on both Hits measures. Compared to the original DSI method,  Hits@1 and Hits@10 improve by 132\% and 46\% for T5-base, and 83\% and 32\% for T5-large. This suggests that the query generation employed in DSI-QG successfully addresses the data distribution mismatch problem that afflicts the original DSI method. 

We further compare our results with those reported by~\citet{wangneural}, who developed the DSI-based technique in parallel with ours. On NQ320k they report their query-generation methods based on a docT5query fine-tuned on the dataset achieves Hits@10 values of 88.48 and 88.45 for the base and large backbones, respectively.  This result supports the superiority of DSI models that exploit query generation, and confirms the backbone model size has minor influence on effectiveness.

Next, we specifically focus on the impact of model size on retrieval effectiveness. We note that the effectiveness of the original DSI method decreases dramatically with a smaller base model. In contrast, model size has relatively little impact on the effectiveness of DSI-QG. This suggests that when using the DSI-QG framework, a large pre-trained T5 model is not necessarily required. The use of a smaller T5 model means that DSI-QG can feature faster retrieval inference time and lower GPU memory requirements.

\subsection{Effectiveness on Cross-lingual Retrieval}
Next we examine the effectiveness of the proposed DSI-QG framework on the cross-lingual retrieval task; recall that these experiments are based on the XOR QA 100k cross-lingual dataset.

In Table~\ref{table2}, we report the results obtained across the different languages. For BM25 + docTquery and DSI-QG, we first generated $n=700$ queries for each document (100 per language) and then separately rank the generated queries for each language using the cross-encoder ranker and the cutoff $m=10$.This resulted in 70 generated queries being used to represent each document (10 for each language). 

The results show that the original DSI model performs much worse on XOR QA 100k than on NQ 320k (cfr. Table~\ref{table1}). In fact, across many languages, DSI-base fails to retrieve any relevant document in the top rank position (Hits@1). This is likely due to the data distribution mismatch problem being further exacerbated by the language gap in the cross-lingual document retrieval task~\cite{Zhang_Zhang_Ao_Gao_Zhuang_Wei_He_2022}. In contrast, our proposed DSI-QG achieves the highest Hits values across all languages with the only exceptions that its Hit@10 on Finnish is lower than that of mDPR, as are the nDCG@10 values for Arab and Russian. 

These results suggest that, with a cross-lingual query generation model, our DSI-QG not only can address the indexing and retrieval gap brought by the data type mismatch but can also address the gap brought by the language mismatch that instead affects the original DSI model.

\begin{figure}[t]
    \centering
    \includegraphics[width=1\linewidth]{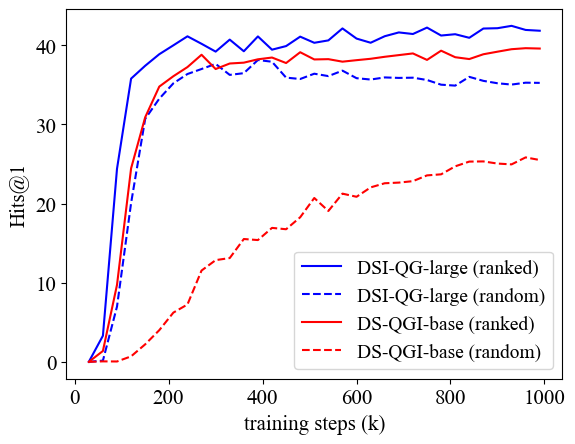}
    \caption{DSI-QG's learning curves on XOR QA 100k. (ranked) denotes that generated queries are ranked by the cross-encoder ranker, (random) denotes that queries are randomly picked from the generated query set. Clearly, ranking and selecting the top-$m$ queries leads to higher effectiveness.}
    \label{fig:rerank}
\end{figure}

\subsection{Impact of Cross-encoder Ranker and Query Selection}
Next, we discuss the effect of different components and factors on our DSI-QG model. Specifically, we study the effect of the cross-encoder ranker and the impact of the rank cut-off $m$ used when ranking and selecting the generated queries.

\subsubsection{Impact of Cross-encoder Ranker}

Figure~\ref{fig:rerank} reports the Hit@1 learning curves on the XOR QA 100k dataset obtained by DSI-QG when trained with and without the cross-encoder ranker. For this experiment, we use the same experimental configuration used for the experiments of Table~\ref{table2}. The plot shows that ranking and selecting the top $m$ generated queries before passing them to the DSI indexing training yields higher Hit@1 convergence than randomly picking $m$ queries from the generated query set. This result is valid for both the base and the large model. This process is however particularly important for the base model to achieve faster convergence and higher final Hits. These results suggest that our use of the cross-encoder ranker, although comes at higher computational and energy costs~\cite{scells2022reduce}, can further improve the effectiveness of DSI-QG by controlling which queries are passed to DSI during indexing.

\subsubsection{Impact of rank cut-off $m$.}
\label{sec_rank_cutoff_results}

\begin{figure}[t]
	\centering
	\includegraphics[width=1\linewidth]{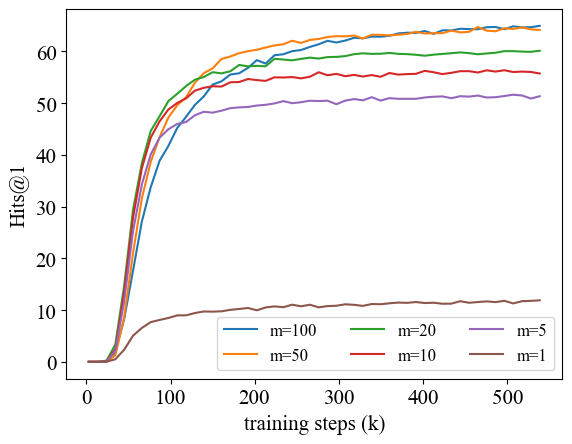}
	\caption{Learning curves of DSI-QG-base trained with different ranking cut-off values $m$ on NQ 320k.}
	\label{fig:num_q}
\end{figure}

 Figure~\ref{fig:num_q} reports the Hit@1 learning curves on NQ 320k for DSI-QG-base trained with different re-ranking cut-off values $m$. For this experiment, we explored cut-off values $m=1, 5, 10, 20, 50, 100$. We note that the value of $m$ also represents the number of queries used to represent each document: when $m=100$, all the $n=100$ generated queries are used and thus the cross-encoder ranker has no effect on the final results. As shown in the plot, effectiveness dramatically increases as $m$ jumps from 1 to 5.
 When $m=5$, DSI-QG already achieves a higher Hits@1 than the original DSI method (reported in Table~\ref{table1}). 
Improvements provided by settings with $m \ge 50$ are not significant compared to values $5 \le m \le 20$. These results align with recent findings in sparse neural retrieval where query generation is adapted as a document expansion method~\cite{lin2021few,mallia2021learning,zhuang2021fast}: a larger number of generated queries can capture more information about the original document, thus providing a better representation of a document. 

 Figure~\ref{fig:num_q} also provides further insights into DSI-QG and its indexing behavior with respect to the number of selected queries for representing a document, $m$. At the beginning of the indexing process, when less than 100,000 training steps $k$ (iterations) have taken place, the setting with $m=100$ is less effective than other settings (with $m>1$). Indeed, it is only when more than $k=300,000$ iterations have taken place, that the setting with $m=100$ achieves the same effectiveness than the setting with $m=50$. Similar behaviors, though less remarked in the figure due to scale, occur when comparing other settings, e.g. $m=50$ against $1<m<50$.

\subsection{Qualitative Analysis of Generated Queries and Ranking}

DSI-QG involves a step of query generation and further ranking and selection of queries to represent a document at indexing. In Table~\ref{example_generation} we report an example of a document from the XOR QA dataset, one of the multilingual query for which this document has been assessed as relevant in the collection, and a sample of the queries that are generated by DSI-QG for the same target language (Russian). The sample of the queries are ordered according to the scores generated for these queries by the cross-encoder ranker. While all the top 3 queries would be used by DSI-QG to represent the document at indexing (when $m>5$), the bottom 3 queries would be discarded by all DSI-QG settings we experimented with in the paper, except when $m=n=100$.

\begin{table}[t]
		\caption{Document, gold query (relevant query for this document as assessed in the dataset), and top 3 and bottom 3 generated queries, ranked according to the cross-encoder ranker used in DSI-QG, for XOR QA's document ``Ryusaku Yanagimoto''. Queries were generated for the Russian language. }
	\begin{small}
		\begin{tabular}{lp{0.35\textwidth}}
			\toprule
			Gold Query & \foreignlanguage{russian}{Как звали первого капитана ``Сорю''?} \\&(What was the name of the first captain of the Soryu?)                                                                                                                                                                                                                                                                                                                                                        \\
			Document   & Ryusaku Yanagimoto\\
			&On 6 October 1941, Yanagimoto was given command of the aircraft carrier \foreignlanguage{russian}{``Sōryū''}, on which he participated in the attack on Pearl Harbor in the opening stages of the Pacific War. He was subsequently at the Battle of Wake Island and the Indian Ocean raids. Yanagimoto chose to go down with his ship when “Soryu” was sunk by United States Navy aircraft at the Battle of Midway. He was posthumously promoted to the rank of rear admiral. \\
			
			\midrule
			Query 1    & \foreignlanguage{russian}{Когда погиб капитан юсуsaka Янагимото?} \\&(When did Captain Yuusaka Yanagimoto die?) \\
			Query 2    & \foreignlanguage{russian}{Как назывался первый корабль, на котором служил японец Рюсаку Янагимото?} \\&(What was the name of the first ship on which the Japanese Ryusaku Yanagimoto served?)  \\
			Query 3    & \foreignlanguage{russian}{В какой войне участвовал Риуsaku Yanagimoto?}\\& (What war did Ryusaku Yanagimoto participate in?) \\
			\midrule
			Query 98   & \foreignlanguage{russian}{Сколько было кораблей на яхте Янагемита?}\\&(How many ships were on Yanagemit’s yacht?) \\
			Query 99   & \foreignlanguage{russian}{Где находилось корабль ``Сорю'' на апрель 2019?}\\&(Where was the Soryu ship in April 2019?) \\
			Query 100  & \foreignlanguage{russian}{Сколько лет правил корабль ``Сорою''?}\\&(How many years did the ship ``Soroy'' rule?) \\              
			\bottomrule                                                                                                                                                
		\end{tabular}
	\end{small}
	\label{example_generation}
\end{table}

\begin{figure}[t]
	\centering
	\includegraphics[width=1\linewidth]{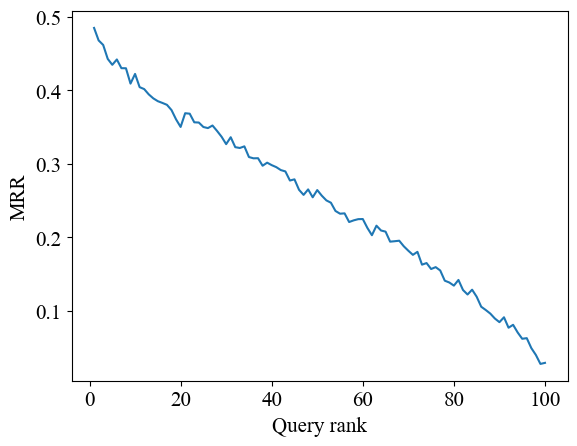}
	\caption{Query effectiveness (MRR) as a function of the rank assigned to each query by the cross-encoder ranker in the DSI-QG pipeline. Document retrieval is performed using the mDPR baseline.}
	\label{fig:query_quality}
\end{figure}

We then generalise the above analysis by considering all queries that have been generated for all documents in the dataset.
Figure~\ref{fig:query_quality} shows the effectiveness, measured in terms of mean reciprocal rank (MRR) of each of the generated queries (ordered by the cross-encoder ranker) at retrieving the relevant document when retrieval is performed using the mDPR baseline. Recall that mDPR is generally highly effective on this dataset, as seen in Table~\ref{table2}(c). Thus, we use mDPR to provide an estimation of ``query quality'' as in this way we decouple this estimation from the training of DSI-QG. The trend observed in the plot suggests that the quality of the generated queries decreases as their rank assigned by the cross-encoder ranker increases, i.e. generally queries in early rank positions are associated to higher mDPR effectiveness than queries at later rank positions, attesting to the importance of ranking queries for use in DSI-QG.

\subsection{Analysis of Length of DSI-QG Input}
While the original DSI uses the full length of a document as input to the Transformer used for indexing the document, DSI-QG uses $m$ queries to represent a document, each of them passed separately as input to the DSI Transformer. We argued that the effectiveness of the original DSI method is limited by the mismatch between the length of the input provided at indexing (documents, usually long) and retrieval (queries, usually short). The new framework we devised, DSI-QG, uses queries for indexing in lieu of documents: this aligns the lengths of the input at indexing (now generated queries, usually short) and the input at retrieval (real queries, usually short). 

We then analyze the input lengths of DSI and DSI-QG  to demonstrate that indeed DSI-QG's indexing lengths are shorter and more aligned with the query lengths observed at retrieval. Input lengths are measured according to the T5 model tokenizer used in DSI, i.e. the number of tokens T5 produces for a text input. 
These statistics are reported in Table~\ref{table3}, and show that indeed for DSI input lengths greatly differ at indexing and retrieval, while these are similar in DSI-QG.

\begin{table}[t]
	\caption{Number of tokens in the DSI input for the original documents (Original), the generated queries (Generated) and the Test Queries used for evaluation at retrieval time. Note that for each document, DSI-QG generates $n=100$ queries: the minimum and maximum lengths for the generated queries then is the min/max of the query lengths averaged for each document.}
	\begin{tabular}{llll}
		\toprule
		Dataset & Input & Mean $\pm$ Std. & [Min, Max]\\
		\midrule
		\multirow{3}{*}{NQ 320k} & Original     & 7,478.03 $\pm$ 8,251.83& {[}3, 153,480{]} \\
		& Generated    & 12.67 $\pm$ 2.05   & {[}8, 29.42{]}   \\
		& Test Queries &  12.07 $\pm$ 3.23      &  {[}7, 32{]}                  \\
		\midrule
		\multirow{3}{*}{XOR QA} & Original     & 164.55  $\pm$ 43.25 & {[}11, 1,640{]}  \\
		& Generated    & 15.10  $\pm$ 1.66  & {[}7, 22.83{]}   \\
		& Test Queries &   14.8  $\pm$  5.55       &  {[}5, 66{]}               \\
		\bottomrule
	\end{tabular}
\label{table3}
\end{table}

\section{Related Work} \label{related}

\subsection{Retrieval via autoregressive generation}
Pretrained transformer-based autoregressive generation models have been shown effective across many NLP tasks~\cite{raffel2019exploring,brown2020language}. Recent studies also explored adapting this type of models to the information retrieval task. 

\citeauthor{cao2021autoregressive} have applied autoregressive generation models to conduct entity retrieval where queries are texts with a mention span and documents are entities in a knowledge base~\cite{cao2021autoregressive,cao2022multilingual}. In this setting, the documents' identifiers are the entity names. 

Different from the entity retrieval setting, \citeauthor{DBLP:journals/corr/abs-2202-06991} proposed the differentiable search index (DSI) scheme~\cite{DBLP:journals/corr/abs-2202-06991}, which is an autoregressive generation model trained to perform ad hoc document retrieval where the input of the model is a natural language query and the model regressively generates documents' identifier strings that are potentially relevant to the given query. 

In another direction, ~\citeauthor{bevilacqua2022autoregressive} proposed the SEAL model which treats ngrams that appear in the collection as document identifiers~\cite{bevilacqua2022autoregressive}. At inference time, SEAL directly generates ngrams which can be used to score and retrieve documents stored in an FM-index~\cite{ferragina2000opportunistic}. 

In contrast to the original DSI scheme and SEAL, our work focuses on augmenting document representations at indexing time so to bridge the gap between indexing and retrieval in the existing autoregressive generation IR systems.

The method of \citet{wangneural}, which was developed in parallel with ours and with no communication between the research groups, shares some commonalities with our DSI-QG in that it also uses query generation for augmenting document representations. However, while their work only considers English document retrieval and thus monolingual augmentation, our work goes beyond that and delves into the setting of cross-lingual document retrieval, offering a more comprehensive approach to the field.

\subsection{Generate-then-rank}
Our DSI-QG indexing framework relies on a cross-encoder model to rank all generated queries in order to identify high-quality queries to represent documents. The intuition behind this design is that, for deep learning models, the generation task is usually a harder task than the classification task. Thus, many deep generation models follow the generate-then-rank paradigm to improve the generation quality. For example, the recent text-to-image generation model DALL$\cdot$E~\cite{ramesh2021zero} also uses a ranker called CLIP~\cite{radford2021learning} to rank all generated images and only present to the users the top-ranked images. On the other hand, while the GPT-3 language model~\cite{brown2020language} has been shown to perform poorly in solving mathematical problems~\cite{hendrycks2021measuring}, ~\citeauthor{cobbe2021training}~\cite{cobbe2021training} have found that simply training verifiers to judge the correctness of the solutions generated by GPT-3 can significantly improve the success of the model for this task. DSI-QG can be seen as following this generate-then-rank paradigm.

\subsection{Query generation for information retrieval}
Our DSI-QG framework relies on the query generation model to generate high-quality and relevant queries to represent each document. Query generation has been the topic of a number of recent works in the field of Information Retrieval.

A common example is docT5query~\cite{nogueira2019doc2query}, a neural document expansion technique that generates relevant queries and appends them to each document in the collection. Then, BM25 is used to perform retrieval on the augmented collection. This simple method can significantly improve on BM25. A follow-up study shows that even completely discarding the original document text and only using the generated queries to represent the documents can achieve better  retrieval effectiveness than using the original document text~\cite{lin2021pretrained}. 

The TILDEv2 model, an effective and efficient sparse neural retrieval model, also uses document expansion based on query generation~\cite{zhuang2021fast}. While one of the query generation methods adopted in TILDEv2 is docT5query, Zhuang\&Zuccon have shown how the TILDE~\cite{zhuang2021tilde} retrieval model can be exploited as a lightweight query generator. The use of TILDE in place of docT5query leads to similar retrieval effectiveness than docT5query but it requires several order of magnitude less computations~\cite{zhuang2021fast,scells2022reduce}. The query generation method we use in DSI-QG is akin to docT5query. While the use of TILDE in place of docT5query for the query generation step of DSI-QG may be attractive because of its lower computational costs, we note that TILDE produces query terms that are independent of each other and thus is unlikely to be effective in for  DSI-QG. In other words: TILDE generates unigram query tokens, not queries (i.e. sequences of tokens) -- and these then are not representative of the inputs that the model will observe at retrieval time (e.g., Table~\ref{table3} shows that queries in the two datasets considered in our work consists of, on average, 12.1 and 14.8 query tokens, respectively). 

Query generation has also been used for the task of domain adaption and for generating training data for the zero-shot setting.  \citeauthor{wang2021gpl} proposed GPL, a framework for training domain adaptation rankers by generating pseudo labels with a query generation model~ \cite{wang2021gpl}. \citeauthor{bonifacio2022inpars} directly used the GPT-3 model~\cite{brown2020language} to generate queries for training rankers in the few-shot setting~\cite{bonifacio2022inpars}. \citeauthor{Luo_Mitra_Gokhale_Baral_2022} introduced a domain-relevant template-based query generation approach which uses a sequence-to-sequence model conditioned on the templates to generate a large number of domain-related queries in a bid to mitigate the train-test overlap issue~\cite{Luo_Mitra_Gokhale_Baral_2022}.

These prior works only focus on the mono-lingual retrieval setting. In contrast, our work also explores the usability of the query generation model for the cross-lingual information retrieval task.

\section{Conclusion} \label{conclusion}
In this paper, we show that the current DSI model is affected by the problem of data distribution mismatch that occurs between the indexing and retrieval phases. This problem negatively impacts the effectiveness of DSI on the mono-lingual passage retrieval task and is further exacerbated in the cross-lingual passage retrieval task, where DSI becomes of impractical use. 

To address this fundamental issue, we propose the DSI-QG indexing framework which adopts a query generation model with a cross-encoder ranker to generate and select a set of relevant queries, which are in turn used to represent each to-be-indexed  document. Our experimental results on both mono-lingual and cross-lingual passage retrieval tasks show that our DSI-QG significantly outperforms the original DSI model and other popular baselines.

\subsubsection*{Acknowledgment}
Part of this work was conducted when Shengyao Zhuang was a research intern at Microsoft STCA, China. 

\bibliographystyle{ACM-Reference-Format}
\bibliography{sigir2023}

\appendix

\end{document}